\documentclass[twocolumn,useAMS,usegraphicx,usenatbib,aas_macros]{mn2e}
\usepackage{array} 
\usepackage{hyperref}
\usepackage{amsfonts}
\usepackage{amssymb} 
\usepackage{amsmath}
\usepackage{graphicx}
\usepackage{enumerate}
\usepackage{mathrsfs}
\usepackage{verbatim} 
\bibpunct{(}{)}{;}{a}{}{,} 

\title[]{Modelling the rotational evolution of solar-like stars: the rotational coupling timescale}
\author[F. Spada et al.]
 {  F.~Spada$^1$, 
  A.~C.~Lanzafame$^{2,3}$,
  A.~F.~Lanza$^3$,
   S.~Messina$^3$ and
  A.~Collier~Cameron$^4$  \\
$^1$ Department of Astronomy, Yale University, 260 Whitney Avenue, New Haven, CT 06511, USA \\
$^2$ Sezione Astrofisica, Dipartimento di Fisica e Astronomia, Universit\`a degli Studi di Catania, Via S. Sofia, 78, 95123, Catania, Italy 
 \\
$^3$ INAF - Osservatorio Astrofisico di Catania, Via S. Sofia, 78, 95123, Catania, Italy \\ 
$^4$ SUPA, School of Physics and Astronomy, University of St Andrews, North Haugh, St Andrews, Fife KY16 9SS, Scotland
}

\begin{document}

\def\aj{{AJ}}                   
\def\araa{{ARA\&A}}             
\def\apj{{ApJ}}                 
\def\apjl{{ApJ}}                
\def\apjs{{ApJS}}               
\def\aap{{A\&A}}                
\def\aapr{{A\&A~Rev.}}          
\def\aaps{{A\&AS}}              
\def\mnras{{MNRAS}}             
\def\nat{{Nature}}              

\date{}

\pagerange{\pageref{firstpage}--\pageref{lastpage}} \pubyear{}

\maketitle

\label{firstpage}

\begin{abstract}
We investigate the rotational evolution of solar-like stars with a focus on the internal angular momentum transport processes.
The double zone model, in which the star's radiative core and convective envelope are assumed to rotate as solid bodies, is used to test simple relationships between the core-envelope coupling timescale, $\tau_c$, and rotational properties, like the envelope angular velocity or the differential rotation at the core-envelope interface.
The trial relationships are tested by fitting the model parameters to available observations via a Monte Carlo Markov Chain method.
The synthetic distributions are tested for compatibility with their observational counterparts by means of the standard Kolmogorov-Smirnov (KS) test.

A power-law dependence of $\tau_c$ on the inner differential rotation leads to a more satisfactory agreement with observations than a two-value prescription for $\tau_c$, which would imply a dichotomy between the initially slow ($P_\mathrm{rot} \gtrsim 3$ d) and fast ($P_\mathrm{rot} \lesssim 3$ d) rotators.
However, we find it impossible to reconcile the high fraction of fast rotators in $\alpha$ Per with the rotation period distributions in stellar systems at earlier and later evolutionary stages. This could be explained by local environmental effects (e.g. early removal of circumstellar discs due to UV radiation and winds from nearby high-mass stars) or by observational biases.
 
The low KS probability that the synthetic and observed distributions are not incompatible, found in some cases, may be due to over-simplified assumptions of the double zone model, but the large relative uncertainties in the age determination of very young clusters and associations are expected to play a relevant role. Other possible limitations and uncertainties are discussed.

\end{abstract}


\section{Introduction}
 
Rotation is an important parameter for solar-like stars. 
It is an essential ingredient of the dynamo process \citep[e.g.][]{Subramanian_Brandenburg:2005}, which in turn powers magnetic activity, it can influence the surface abundance of light elements through rotationally-induced mixing \citep{Zahn:1993}, and it is related to the formation and evolution of planetary systems \citep[see, e.g.,][]{Chang_ea:2010, Lanza:2010}.

Rotation period measurements in open clusters and young associations provide a valuable source of information on angular momentum evolution of late-type stars. 
In recent years,  photometric measurements of the stellar rotation period, based on the rotational modulation of the light curve induced by photospheric starspots, have replaced the measurements based on the $v\, \sin i$ as  derived from spectral line profiles. 
This has considerably improved both the number  and the precision of available rotation periods.  
Such observational constraint, however, has not been exploited in its full potential yet. 
In fact, most of the analyses carried out so far have considered only limiting cases such as the subsets of the faster and slower rotators at given ages.

During the pre-main sequence (PMS) phase, solar-mass stars undergo a global contraction. 
However, for the first few Myr, the interaction with a circumstellar disc drains angular momentum from the star, thus delaying its spin up for the (variable) duration of the disc lifetime.
This process is  not understood in detail yet (see, e.g., \citealt{Cameron_Campbell:1993, Shu_ea:1994, Matt_ea:2010}) and is usually modelled by means of very simplified assumptions, e.g., the disc-locking hypothesis \citep[][]{Koenigl:1991}.
Once on the main sequence (MS), the rotational evolution is driven by the loss of angular momentum through a magnetised stellar wind \citep{Weber_Davis:1967,Kawaler:1988,Chaboyer:1995a,Chaboyer:1995b}.

Surface period measurements are also an indirect probe of the internal rotation profile of these stars. 
Phenomenological modelling of MS angular momentum evolution \citep[the so-called double zone model, DZM in the following:][see also Sect. \ref{sec:dzmodel}]{MacGregor_Brenner:1991, Keppens_ea:1995, Allain:1998, Bouvier:2008b}, assuming that the radiative core and the convective envelope rotate rigidly at all ages, has successfully reproduced the observations provided a certain amount of differential rotation between the core and the envelope is allowed. 
This is parameterised through a coupling timescale $\tau_{\rm c}$, which determines the rate of angular momentum exchange between the two regions.
From the comparison of synthetic rotational tracks and the upper and lower percentiles of the observed period distributions, \citet{Bouvier:2008b} concluded that $\tau_{\rm c}$ is of the order of $10$ Myr for the stars that begun their evolution on the ZAMS as fast rotators (i.e., with an initial rotation period of $\sim$ 1~d), while it is remarkably longer ($\tau_{\rm c} \sim 100$ Myr) for initially slow rotators (i.e., with initial periods $\sim$ 10~d). 
A possible connection between this dichotomy and the presence or absence of a planetary system orbiting the star is further discussed in \citealt{Eggenberger_ea:2010}. 
An understanding of the physical processes that give rise to such a dichotomy and eventually establish a nearly solid-body rotation within the age of the Sun \citep[as observed through helioseismology, see e.g.][]{Thompson_ea:2003} is still lacking. 

Our aim is twofold: on the one hand, we try to extract the whole statistical information from the observed rotation period distributions. 
In recent years,  rotation period measurements have improved significantly both in number and in precision.  
In spite of the age spread within cluster members, uncertainties on cluster age estimates and other systematic effects, good sample completeness and homogeneity has already been achieved.
We exploit this information comparing the observed period distributions with synthetic ones, constructed by  evolving the Orion Nebula Cluster (hereafter ONC) distribution by means of the DZM. 
The modelling of the period distributions based on the DZM requires some prescription on the dependence of $\tau_c$ on the angular velocity, to comply with the findings of \citet{Bouvier:2008b} mentioned above.  
On the basis of intuitive physical arguments, we formulate some hypotheses on the scaling of $\tau_c$.
The set of parameters in the DZM that produces the best fit of the synthetic distributions to the observed ones is determined by means of an iterative procedure based on a Markov Chain Monte Carlo (MCMC) sampling.
This suffices to the purpose of comparing the different hypotheses on $\tau_c$ and to test the overall capability of the DZM to satisfy the latest observational constraints.

A direct comparison of theoretical modelling of stellar rotation with observed period distributions has already been attempted in the past by, e.g., \citet{Tinker_ea:2002}.
They evolved the ONC period distribution to the age of the Pleiades and the Hyades by means of the Yale Rotating Evolution Code \citep{Guenther_ea:1992}, mainly aiming at determining the parameters in the wind braking law.
The present study is novel in many respects, because we rely upon a simpler modelling, focussing on the coupling timescale. Moreover, a much larger sample of observed period distributions is used.

\section{Rotational period data}
\label{sec:obs}

We use the rotation period distributions extracted from photometric surveys of stellar open clusters and young associations, as available in the literature (see Table \ref{tab:data} for the references).

The rotational data for the associations TW Hydrae, $\beta$ Pictoris, Tucana--Horologium, Columba, Carina (age $\lesssim 30$ Myr), AB Doradus (age $\sim 120$ Myr) are taken from \cite{Messina_ea:2010a} and for the cluster M 11 (age $\sim 230$ Myr) from \cite{Messina_ea:2010b} and are used for the first time for this kind of investigation. 
Remarkably, these young associations fill an age gap previously present  between Tau-Aur ($\sim 6$ Myr old) and $\alpha$ Per ($\sim 70$ Myr) in the mass range $0.9 \leq M_*/M_\odot \leq 1.1$.

The ages reported here are the most recent literature estimates. 
In particular, the association AB Dor is considered almost coeval with the Pleiades, according to the analysis of \citet{Messina_ea:2010b}, who revised a previously proposed age of $70$ Myr.
A remarkably extensive rotational dataset for the Pleiades have been produced by \cite{Hartman_ea:2010}.

The rotational evolution significantly depends on stellar mass, owing to different durations of the PMS phase, convective envelope depths, etc.
As we are primarily interested in the rotational evolution of solar analogues, a subsample of each period distribution in the mass range $0.9 \leq M_*/M_\odot \leq 1.1$ was selected.
For very young clusters (i.e. age $\leq 10$ Myr), which are affected by a substantial reddening, the effective temperature of the stars was determined on the basis of their spectral type as given by \citet{Rebull_ea:2004}.
For MS objects, the dereddened $(B-V)$ and $(V-I)$ colours were employed for the same purpose. 
In both cases, the determination of the mass from $T_\mathrm{eff}$ or the colour index was performed by direct comparison with the theoretical isochrones of \cite{Baraffe_ea:1998}. 

\begin{table}
\caption{ Names, estimated ages and number of stars in the mass range $0.9 \leq M_*/M_\odot \leq 1.1$ for the stellar systems used in this work. Stellar associations are in bold.}
\label{tab:data}
\begin{center}
\begin{tabular}{c|c|c|c}
\hline
\hline
Object name & Age [Myr] & \# stars & Reference \\
\hline
ONC & 2 & 57 & \ref{reb} \\
NGC 2264 & 4 & 52 & \ref{reb} \\
NGC 2362 & 5 & 43& \ref{ir08} \\
{\bf Tau-Aur} & 6 & 25& \ref{reb} \\
{\bf TW Hya}, {\bf $\beta$ Pic}  & 10 & 12& \ref{smass} \\
{\bf Tuc-Hor}, {\bf Col}, {\bf Car}  & 30 & 46 & \ref{smass} \\
$\alpha$ Per & 70 & 43 & \ref{smcomp} \\
{\bf AB Dor}, Pleiades & 120 & 109 & \ref{smass},  \ref{har10} \\
M 50 & 130 & 62 & \ref{ir09} \\
M 35 & 150 & 88& \ref{mei} \\
M 34 & 220 & 17& \ref{ir06} \\
M 11 & 230 & 15 & \ref{sm10} \\
M 37 & 550 & 89 & \ref{h09} \\
\hline
\end{tabular}
\end{center}
\medskip
\protect \begin{enumerate}[(a)]
\item \label{reb}: compilation, see \protect \cite{Rebull_ea:2004} and references therein;
\item:  \label{ir08} \protect \cite{Irwin_ea:2008}; 
\item: \label{smass} \protect \cite{Messina_ea:2010a}; 
\item: \label{smcomp} compilation, see \protect \cite{Messina_ea:2001}, \protect \cite{Messina_ea:2003} and references therein;  
\item: \label{har10} \protect \cite{Hartman_ea:2010}; 
\item: \label{ir09} \protect \cite{Irwin_ea:2009}; 
\item: \label{mei} \protect \cite{Meibom_ea:2009}; 
\item: \label{ir06} \protect \cite{Irwin_ea:2006}; 
\item: \label{sm10} \protect \cite{Messina_ea:2010b}; 
\item: \label{h09} \protect \cite{Hartman_ea:2009}.

\end{enumerate}
\end{table}

\begin{figure}
\begin{center}
\includegraphics[width=0.45\textwidth]{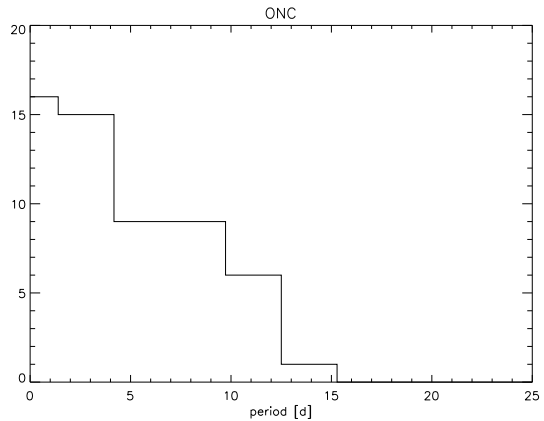}
\caption{Observed period distribution for the $57$ stars of ONC in the selected mass range.}
\label{fig:onc}
\end{center}
\end{figure}

The ONC period distribution, providing the initial conditions for our rotational modelling, is shown in Figure \ref{fig:onc}.
In the restricted mass range which is of interest to us, the distribution is unimodal, with a fairly pronounced peak of fast rotators (i.e., with $P_\mathrm{rot}\leq 4$~days).

\section{Numerical procedure}
\label{sec:method}

We construct a series of synthetic rotation period distributions at the ages reported in Table \ref{tab:data}.
Each period in the ONC solar analogue subsample is evolved forwards in time with the DZM, calculated for a $1~M_\odot$ star. 
The synthetic distributions are tested for compatibility with their observational counterparts by means of the standard Kolmogorov-Smirnov (KS) test \citep[see, e.g., sect. 14.3 of][]{NR:1992}.
The input to our numerical procedure is an initial guess for the parameters appearing in the equations \eqref{eq:dzm} of the DZM.
The parameter space is then sampled by means of the MCMC method, which performs random jumps in each parameter, guided by a probabilistic acceptance rule (see Sect. \ref{sec:mcmc}).
The best-fitting set of model parameters is obtained in output.

\subsection{Parameters in the DZM}
\label{sec:dzmodel}
Our simple model for the rotational evolution of a solar-like star assumes that the radiative core and the convective envelope both rotate as solid bodies \citep{MacGregor_Brenner:1991,Keppens_ea:1995,Allain:1998}. 
The envelope is expected to be strongly coupled by the very efficient angular momentum redistribution associated with turbulent convection.
On the other hand, a large scale magnetic field, even as weak as $\lesssim 1$~G, can maintain a condition of rigid rotation in most of the core (see, e.g. \citealt{Spada_ea:2010}).

At any given time $t$, the angular momentum of the core, $J_c$, and of the envelope, $J_e$, are:
\begin{eqnarray*}
J_c(t) = I_c(t)\ \Omega_c(t) \ \ \ ; \ \ \ J_e(t) = I_e(t)\ \Omega_e(t) ,
\end{eqnarray*}
where $I_c$, $I_e$ and $\Omega_c$, $\Omega_e$ are the moments of inertia and the angular velocities of the core and the envelope, respectively.
Our evolution begins at the ONC age, i.e. $t_0 \sim 2$ Myr. 
Stars are fully convective until $t_0$, thus justifying a solid-body rotation as our initial condition, i.e. $\Omega_e(t_0) = \Omega_c(t_0)$.

The evolution of the angular momenta is governed by two coupled differential equations: 
\begin{eqnarray}
\nonumber
\frac{dJ_c}{dt} &=& -\frac{\Delta J}{\tau_c} + \left( \frac{2}{3}~R_c^2~\frac{dM_c}{dt}  \right) \Omega_e ;   \\
\label{eq:dzm}
\\
\nonumber
\frac{dJ_e}{dt} &=& +\frac{\Delta J}{\tau_c} - \left( \frac{2}{3}~R_c^2~\frac{dM_c}{dt}  \right) \Omega_e - \left. \frac{dJ_e}{dt} \right|_\mathrm{wind} ,
\end{eqnarray}
where $R_c$ and $M_c$ are the radius and mass of the radiative core.
The stellar structure quantities and their time derivatives are taken from PMS and MS evolutionary models by \citet{Baraffe_ea:1998}.
In equations \eqref{eq:dzm}, the core and the envelope exchange an amount of angular momentum $\Delta J \equiv \dfrac{I_e\ J_c - I_c\ J_e}{I_e+I_c}$ at a rate determined by the coupling timescale $\tau_c$.
During the PMS, the  growth of the core at the expenses of the envelope produces an angular momentum transfer, accounted for by the second term in the right-hand side of both equations \eqref{eq:dzm}.
Because of the magnetised wind, the envelope loses angular momentum at a rate $ \left. \dfrac{dJ_e}{dt} \right|_\mathrm{wind}$, for which the following parametric formula is used \citep{Kawaler:1988}:
\begin{equation}
\label{eq:wind}
 \left. \frac{dJ_e}{dt} \right|_\mathrm{wind} = K_w \left( \frac{R_*/R_\odot}{M_*/M_\odot} \right)^{1/2} \mathrm{min}(\Omega_e^3,\Omega_\mathrm{sat}^2\ \Omega_e) ,
\end{equation}
where $K_w$ and $\Omega_\mathrm{sat}$ determine the braking intensity and saturation threshold, respectively, while $R_*/R_\odot$ and $M_*/M_\odot$ are the stellar radius and mass in solar units.
Finally, the angular momentum loss due to the interaction with the disc during early PMS evolution is taken into account with the disc-locking hypothesis \citep{Koenigl:1991}, i.e.:
\begin{equation*}
\frac{d \Omega_e}{dt} = 0 \ \ \mathrm{while}\ \ \ t \leq \tau_\mathrm{disc} .
\end{equation*}
In other words, the net effect of the interaction with the disc is that of keeping the surface angular velocity of the star constant for the whole disc lifetime.

The solutions of equations \eqref{eq:dzm} depend on the parameters $K_w$, $\Omega_\mathrm{sat}$, $\tau_c$ and $\tau_\mathrm{disc}$. 
In applying the DZM to a whole distribution of rotators, the wind law parameters  $K_w$ and $\Omega_\mathrm{sat}$ can be assumed to be the same for all the stars; for $\tau_c$ and $\tau_\mathrm{disc}$, on the contrary, such an assumption is rather crude.
It has already been established that it is impossible to reconcile the lower and upper envelopes of the rotation period distributions with the same value of $\tau_c$ \citep[e.g.][]{Bouvier:2008b}.
This is dealt with by simple assumptions on the dependence of $\tau_c$ on stellar rotation, as discussed below (see Sect. \ref{sec:tauc}).
A realistic treatment of the star-disc interaction is a complex task. 
Even under the disc-locking assumption, the value of $\tau_\mathrm{disc}$ for each star in the sample should be extracted from a distribution.

Although information on the distribution of disc lifetime, as well as on its dependence on the presence of high mass stars in the neighbourhood, is becoming increasingly available \citep[e.g.][]{Hernandez_ea:2008, Kennedy_Kenyon:2009}, we refrain from exploiting it in full details so to keep our model as simple as possible.

\subsection{Hypotheses on the coupling timescale}
\label{sec:tauc}

We consider the following prescriptions on the dependence of $\tau_c$ on stellar rotation:
\begin{itemize}
\item A two-valued function of the form:

\begin{equation}
\label{eq:2val}
\tau_c = \left\{ 
\begin{array}{cc}
 10 \ \mathrm{Myr} & \mathrm{if} \ \ \Omega_e(t_0) \geq \Omega_\mathrm{crit}      \\
 \tau_0 & \mathrm{otherwise}   
\end{array}
\right. .
\end{equation}
This is the simplest implementation of the early empirical finding (e.g. \citealt{Krishnamurthi_ea:1997}, see also \citealt{Bouvier:2008b}) that fast rotators are well described by a nearly solid-body inner profile, while a sizeable decoupling is necessary for slow rotators.
The value of $10$ Myr used for initially fast rotators is so short as to ensure nearly instantaneous coupling at all ages \citep[see also][]{Denissenkov_ea:2010}.
The timescale for slow rotators, $\tau_0$, and the initial angular velocity threshold, $\Omega_\mathrm{crit}$, are treated as free parameters to be determined by means of the MCMC procedure.

\item A power law dependence on the instantaneous amount of differential rotation:
\begin{equation}
\label{eq:powl}
\tau_c(t) = \tau_0 ~ \left[ \frac{\Delta \Omega_\odot}{ \Delta \Omega(t)} \right]^{\alpha} \ ,
\end{equation}
where $\Delta \Omega(t) = \Omega_c(t) - \Omega_e(t)$ and $\Delta \Omega_\odot$ is the inner differential rotation of the present Sun, assumed as a reference value (we used $\Delta \Omega_\odot = 0.2 \ \Omega_\odot$).

Note that this choice produces a time-dependent $\tau_c$. 
This is an attempt to account for  the effect of an enhanced angular momentum transport due to turbulent viscosity.  
The viscosity enhancement could be  the result of either a rotational mixing \citep[e.g.][]{Zahn:1993} or of hydromagnetic instabilities \citep{Spruit:1999,Spruit:2002}  powered by the available kinetic energy stored in the differential rotation.
The form of equation \eqref{eq:powl}, with $\alpha>0$, is chosen to produce a shorter coupling timescale (which mimics a greater effective viscosity, ensuring a stronger coupling) when a greater differential rotation is present.
$\tau_0$ and $\alpha$ are regarded as free parameters.
\end{itemize}

Other simple prescriptions, e.g. a dependence of $\tau_{c}$ on the surface angular velocity $\Omega_e(t)$, failed to give suitable results and are not discussed in the following.

\subsection{The MCMC method}
\label{sec:mcmc}

Our application of the MCMC method \citep[see, e.g.][]{Tegmark_ea:2004} constructs a sequence of values for each parameter of the DZM, approaching the best fitting region of the parameter space quite rapidly (burn-in phase, usually a few hundreds of steps), and then sampling this restricted neighbourhood by performing a random walk within it.

The routine is initialised providing a starting guess for the parameters.
At the beginning of each step, the current set of parameters is used to evolve forwards in time the ONC period distribution to each of the ages in Table \ref{tab:data} by means of the DZM. 
A KS test is then performed comparing the synthetic period distributions with their observational counterparts. 
Calling $\bar P_{\mathrm{KS},i}$ the probability of the synthetic and observed distributions to be significantly different for a certain cluster $i$ and viewing the individual KS tests as independent from each other, we define the quantity $Q$ by adding these probabilities together:
\begin{equation*}
Q = \sum_{i=1}^N \bar{P}_{\mathrm{KS},i} \ ,
\end{equation*}
with $N$ being the number of rotation period distributions used.
Two additional constraints come from the rotational properties of the present Sun, namely: 
\begin{equation}
\label{eq:sun_con}
\Omega_e(t_\odot) = \Omega_\odot \ \ \ ; \ \ \ \frac{\Omega_c(t_\odot)-\Omega_e(t_\odot)}{\Omega_e(t_\odot)} \leq 0.2 ,
\end{equation}
which are taken into account by suitable additive terms in $Q$.
The MCMC procedure is used to iterate the model parameters towards the minimum of $Q$, which corresponds to the best fit.

After an evaluation of $Q$, the following step begins with a tentative random jump in each parameter. 
The new set of parameters just generated is retained in the chain or rejected following the stochastic Metropolis-Hastings rule \citep{Metropolis_ea:1953,Hastings:1970}:   
\begin{eqnarray*}
\Delta Q < 0  &:& \ \mathrm{accept} ; \\
\Delta Q > 0  &:& \ \mathrm{accept } \ \mathrm{with} \ \mathrm{probability} \propto e^{-\Delta Q} ,
\end{eqnarray*}
where $\Delta Q$ is the difference between the current value of $Q$ and its value at the last accepted step.
This rule ensures a quick descent towards the global minimum, with a non-zero probability of escaping from possible local minima encountered during the path.

\begin{table}
\caption{  Best fitting parameters estimated applying the MCMC method. 
The units of $K_w$ are g~cm$^2$~s.
}
\begin{center}
\begin{tabular}{c|c}
\hline
\hline
two-valued $\tau_c$ & power law $\tau_c$  \\
(cf. equation \ref{eq:2val}) &  (cf. equation \ref{eq:powl})  \\
\hline
$\left< K_w \right>=(5.40 \pm 0.054) \cdot 10^{47}$  & $ \left<K_w \right>=(5.97 \pm 0.13) \cdot 10^{47}$ \\
\\
$\left< \tau_0 \right>= 128 \pm 3.84$ Myr & $\left< \tau_0 \right>= 57.7 \pm 5.24$ Myr \\
\\
$\left< \Omega_\mathrm{crit} \right>= 3.89 \pm 0.14 \ \Omega_\odot$ & $ \left<\alpha \right>=0.076 \pm 0.02$  \\
\hline
\end{tabular}
\end{center}
\label{tab:results}
\end{table}

\section{Results}
\label{sec:results}

The procedure described above was applied to determine the best values of $K_w$, $\tau_0$ and $\Omega_\mathrm{crit}$, in equation \eqref{eq:2val}, or $\alpha$, in the case of equation \eqref{eq:powl}, respectively. 
The parameters $\Omega_\mathrm{sat}$ and $\tau_\mathrm{disc}$ were held fixed.
We used the following fiducial values:
\begin{eqnarray*}
\Omega_\mathrm{sat} = 5.5 \, \Omega_\odot \ \ \ ; \ \ \  \tau_\mathrm{disc} = 5.8 \, \mathrm{Myr}.
\end{eqnarray*}
These choices are motivated by independent observational estimates, namely the angular velocity saturation threshold for the X-ray emission of late-type stars \citep[see figure $3$ of][]{Pizzolato_ea:2003} for $\Omega_\mathrm{sat}$, and the age at which about $10\, \%$ of stars still possess a circumstellar disc \citep[see figure $1$ of][]{Mamajek:2009} for $\tau_\mathrm{disc}$. 
This is also in good quantitative agreement with earlier works \citep[e.g.][]{Bouvier:2008b, Irwin_Bouvier:2009}. 

The best-fitting values of $K_w$, $\tau_0$ and $\Omega_\mathrm{crit}$ were calculated from the portion of the MCMC chain that effectively sampled the region of the parameter space around the minimum (in both cases, we had $\gtrsim 10^4$ steps).
For each parameter, we used the mean value and the standard deviation as estimates of the best fitting value and its uncertainty, respectively. 
The results are summarised in Table \ref{tab:results}. 
Note that the value of $K_w$ found, of the order of $10^{47}$ g cm$^2$ s, is in good agreement with previous works (see, e.g. \citealt{Kawaler:1988}, \citealt{Allain:1998}).

\begin{figure}
\begin{center}
\includegraphics[width=0.45\textwidth]{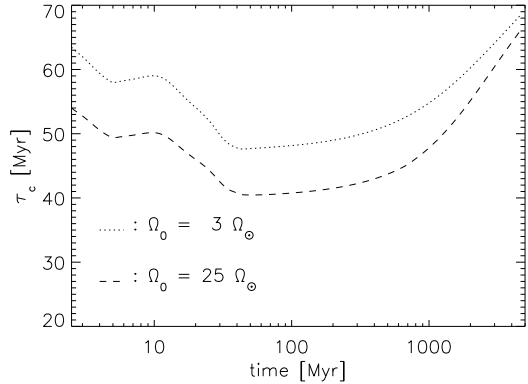}
\caption{Values of $\tau_c$ determined by the power-law prescription with the best-fitting $\tau_0$ and $\alpha$ (see Table \ref{tab:results}) for two initial velocities representative of slow (dotted line) and fast (dashed line) rotators, respectively.}
\label{fig:tauc_plaw}
\end{center}
\end{figure}

To allow the comparison of the two different prescription for $\tau_c$, we show in Figure \ref{fig:tauc_plaw} the evolution of $\tau_c$ in the power-law case for two initial values of $\Omega_e$ representative of slow and fast rotators.
Remarkably, the best-fitting parameters produce a $\tau_c \gtrsim 40$ Myr for fast rotators, significantly longer than $10$ Myr, which was used in the two-valued case to ensure nearly solid-body rotation at all ages, as suggested by previous works. 

\begin{figure*}
\begin{center}
\includegraphics[width=0.8\textwidth]{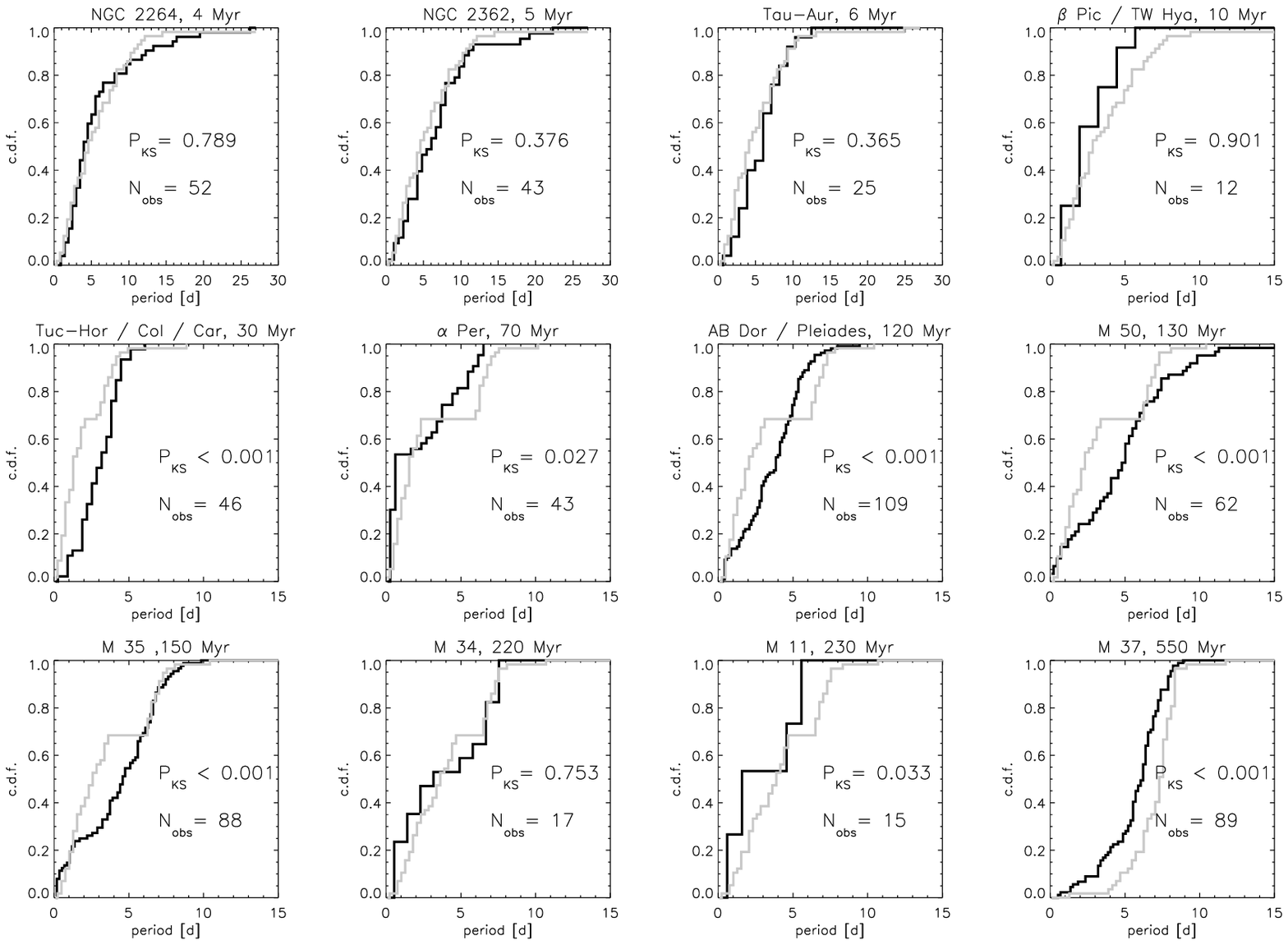}
\caption{Synthetic (grey line) and observed (black line) rotation period cumulative distributions for the two-valued $\tau_c$ case. The complementary probability of being drawn from significantly different distributions, as calculated by means of the KS test, is reported in each panel together with the number of objects in the observed distribution.}
\label{fig:pks_2val}
\includegraphics[width=0.8\textwidth]{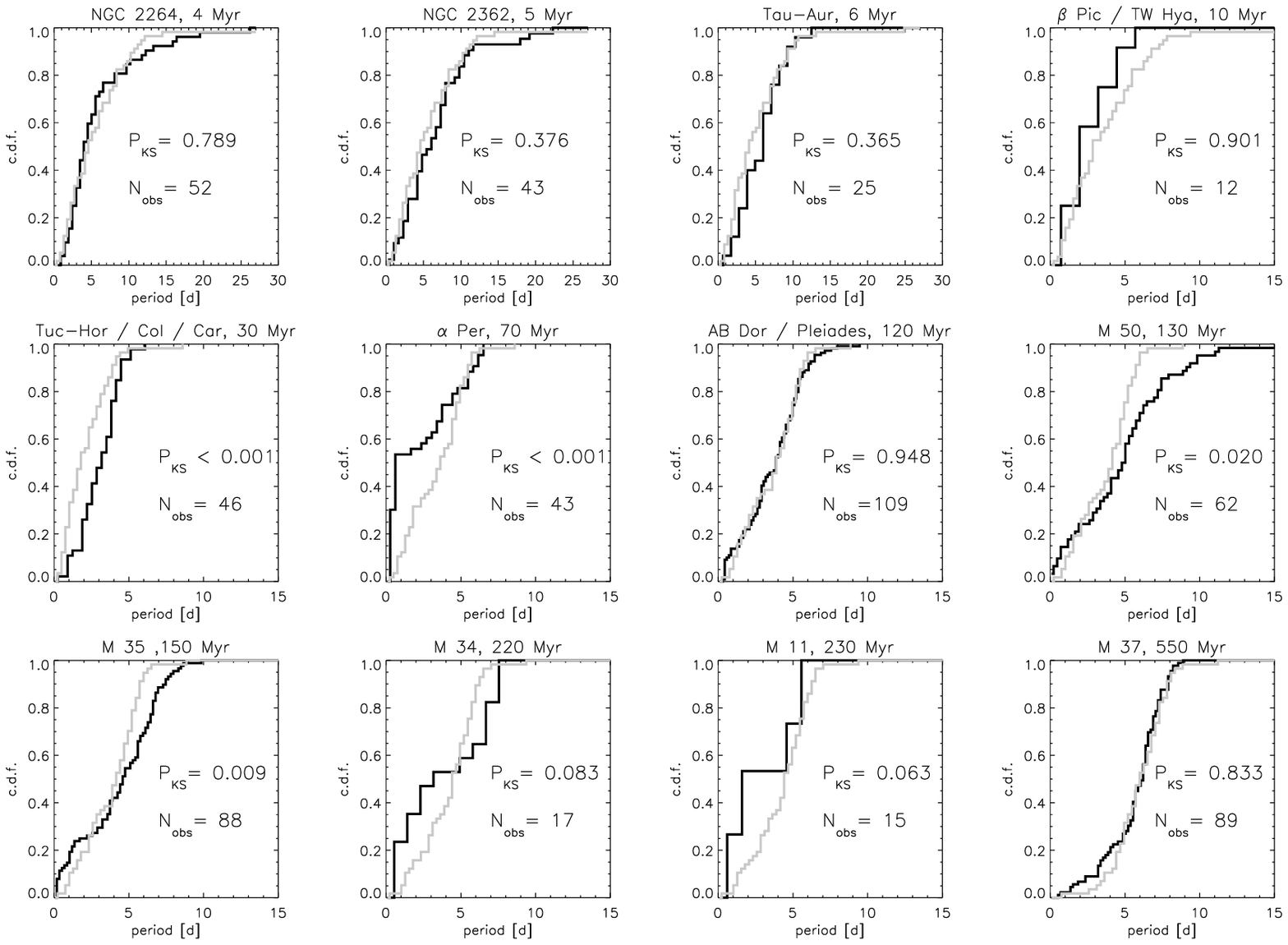}
\caption{Synthetic (grey line) and observed (black line) rotation period cumulative distributions for the power law $\tau_c$ case. The complementary probability of being drawn from significantly different distribution, as calculated by means of the KS test, is reported in each panel together with the number of objects in the observed distribution.}
\label{fig:pks_plaw}
\end{center}
\end{figure*}

A comparison of the quality of the fitting in the two cases, based on the individual KS tests, is presented in Figures \ref{fig:pks_2val} and \ref{fig:pks_plaw}, which show the synthetic and observed cumulative distributions (CDF) for the various clusters.
The  KS compatibility probability, $P_{\mathrm{KS},i}$, i.e., the probability of the two distributions for cluster $i$ of being not incompatible, is reported in each panel along with the number of stars in the respective observed distribution (note that $P_{\mathrm{KS},i} = 1 - \bar P_{\mathrm{KS},i}$).
The smaller the value of $P_{\mathrm{KS},i}$, the worse the fitting between the calculated and the observed distributions for cluster $i$. 

The photometric periods on which our analysis is based are Fourier-derived from the rotational light modulation induced by starspots. The amplitude of such a modulation falls below the minimum value currently detectable from the ground ($\simeq 0.01$ mag) for MS stars with $P_\mathrm{rot} \gtrsim 10$ d (see also Sect. \ref{sec:disc}).
As a consequence, the photometric rotation period of slow rotators is hardly derived and the observed distributions may suffer a lack of completeness beyond this value of the rotation period. 
To account for this effect, periods in the synthetic distribution greater than the maximum of the corresponding observed distribution are not considered when calculating $ \bar P_{\mathrm{KS},i}$ and $P_{\mathrm{KS},i}$.

A good agreement ($P_{\mathrm{KS},i} \gtrsim 0.37$) is achieved for the first $\sim 10$ Myr in both Figures.
The first two panels refer to clusters younger than the assumed $\tau_\mathrm{disc}$; as a consequence, the corresponding synthetic distributions show no evolution with respect to ONC, by virtue of the disc-locking constraint.
For ages greater than $\tau_\mathrm{disc}$ but lower than $\sim 30$~Myr, the spin-up dominates due to the stellar contraction. 
At $6$ and $10$ Myr, therefore, the agreement with observations mainly rely on $\tau_\mathrm{disc}$ and on the stellar structure and evolution model, while it is quite insensitive to $\tau_c$ and $K_w$. 
For ages older than approximately 30~Myr, the agreement is strongly dependent on $K_w$ and on the model assumed for $\tau_c$. For the two-value $\tau_c$ model, the only satisfactory agreement is found for M~34 (220~Myr). 
For the power-law $\tau_c$ model, a satisfactory agreement is found for the Pleiades (120~Myr) and M~37 (550~Myr). 
We note, however, that the CDFs derived from the observations do not follow an evolution as regular as expected from the models. 
The $\alpha$ Persei CDF, in particular, displays a sharp rise at short rotation periods due to an excess of  fast and ultra-fast rotators. 
Such a sharp rise is not seen in any other CDFs, except, perhaps, in M~34 and M~11, although the number of periods available is too low to draw any definitive conclusions. 
The two-value $\tau_c$ model can account for this sharp rise, but then fails to reproduce the smooth CDF shape of AB Dor/Pleiades, M~50, etc. 
The power-law $\tau_c$ model, on the other hand, cannot reproduce any sharp rise in the CDF and therefore fails to reproduce the $\alpha$ Persei, M~34 and M~11 CDFs. 

\begin{figure*}
\begin{center}
\includegraphics[width=0.8\textwidth]{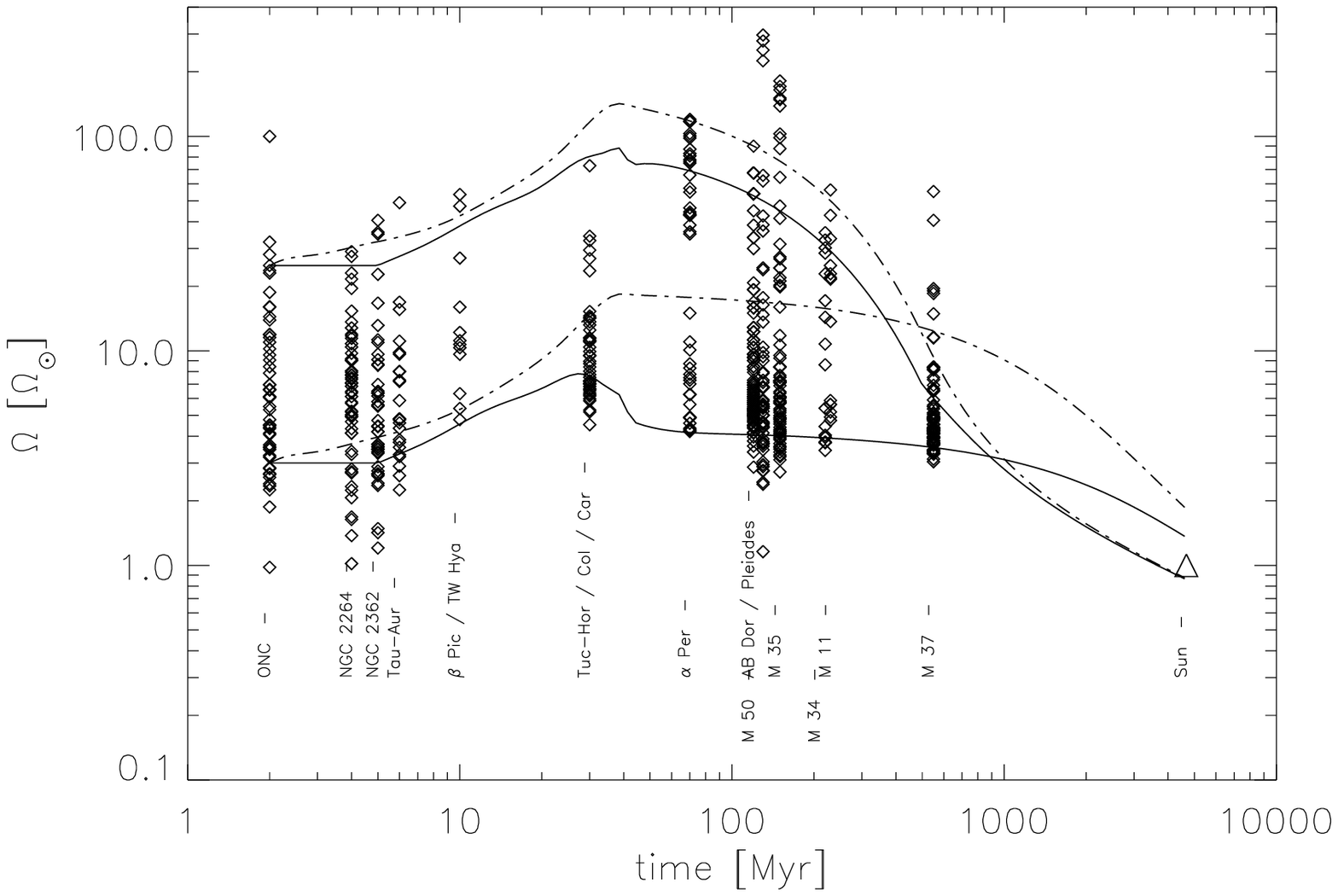}
\caption{Evolution of two rotators, according to the DZM as modified by the two-valued $\tau_c$ prescription in equation \eqref{eq:2val} and using the parameters on the left of Table \ref{tab:results}, with initial conditions representative of the fast and slow groups ($\Omega_e(t)$ are plotted as solid lines and $\Omega_c (t)$ as dash-dotted lines). Observed rotation period distributions (diamonds) and the Sun (triangle) are also plotted for comparison.} 
\label{fig:ev_2val}
\includegraphics[width=0.8\textwidth]{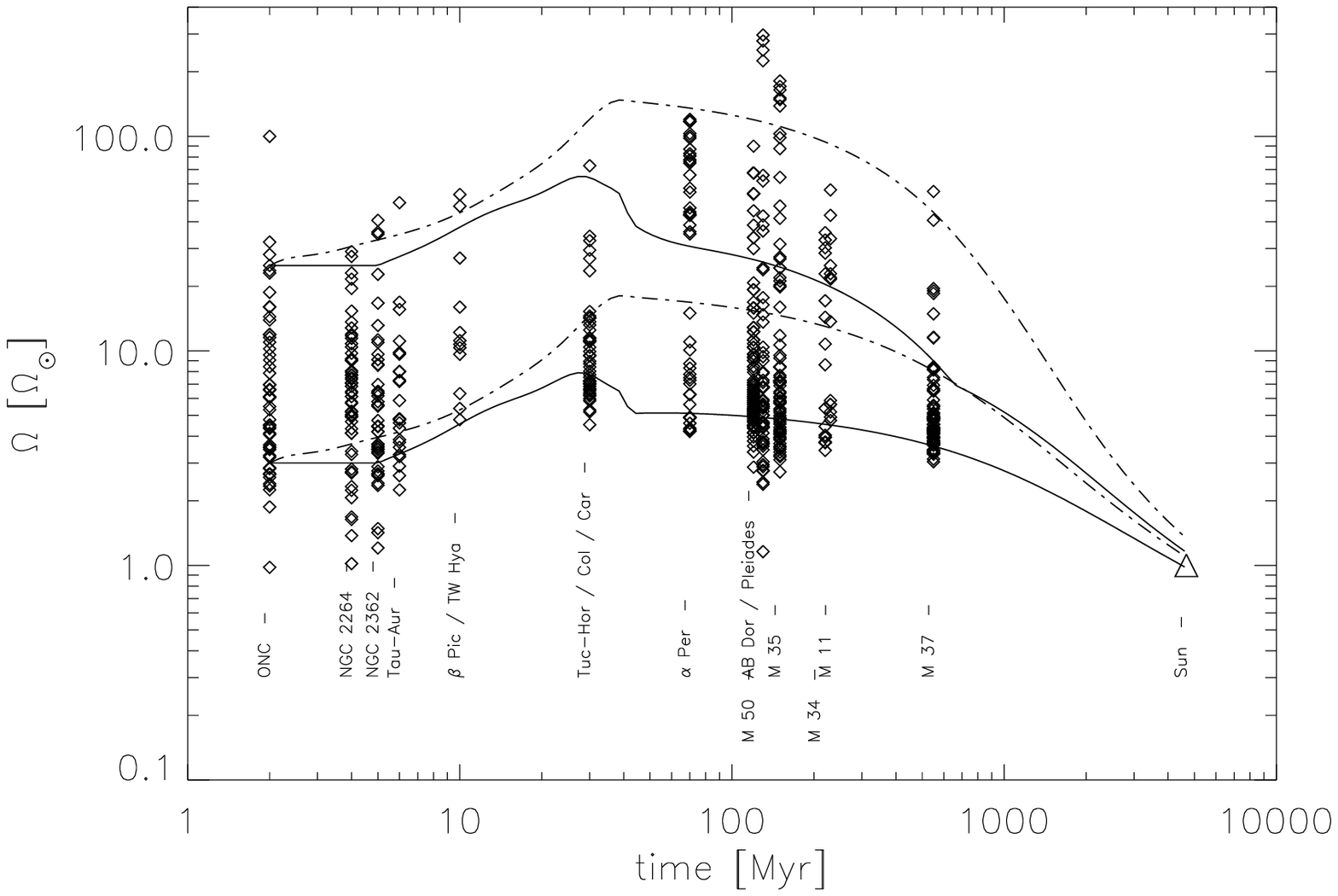}
\caption{Evolution of two rotators, according to the DZM as modified by the power law $\tau_c$ prescription in equation \eqref{eq:powl} and using the parameters on the right of Table \ref{tab:results}, with initial conditions representative of the fast and slow groups ($\Omega_e(t)$ are plotted as solid lines and $\Omega_c (t)$ as dash-dotted lines). Observed rotation period distributions (diamonds) and the Sun (triangle) are also plotted for comparison.}
\label{fig:ev_plaw}
\end{center}
\end{figure*}

Without any angular momentum transfer from the core to the envelope, the maximum angular velocity is expected at about the Tuc-Hor / Col / Car age, i.e., $\approx 30$~Myr (see also Figures \,\ref{fig:ev_2val} and \ref{fig:ev_plaw}).
However, this is not observed because the evolution from the  Tuc-Hor / Col / Car age (30 Myr) to $\alpha$ Persei (70 Myr) appears consistent with a continuation of the spin-up after the ZAMS.
The spin-up of fast and ultra-fast rotators from the Tuc-Hor / Col / Car age (30 Myr) to the $\alpha$ Persei (70 Myr) age seems even faster than for slower rotators.
In the DZM framework, such a behaviour cannot be reproduced unless, for the fast and ultra-fast rotators, a remarkable amount of angular momentum is transferred from the core to the envelope within a very short timescale ($\sim 30-40$~Myr).
In this case, however, very little angular momentum  would be left in the core for the later evolution.

After the peak in the angular velocity, a generally monotonic spin-down is expected.
The fast rotator excess in M~34 (220 Myr) and M~11 (230~Myr), however, hampers the possibility of obtaining a satisfactory fit at all ages. 
In fact, the percentage of observed fast rotators ($P_{\rm rot}\sim$ 1 - 2 d) tends to decrease from $\alpha$ Per to M~35, but then increases abruptly in M~34 and M~11.
Such a distribution evolution is clearly inconsistent with the DZM alone.  

Despite such limitations, which will be discussed in some more details in Sect.\,\ref{sec:disc}, the power-law $\tau_c$ model produces a more satisfactory fit to the data, particularly since it reproduces the shape of most of the observed CDFs also giving a high $P_{\mathrm{KS},i}$ value for AB Dor / Pleaides and M~37.
It requires, however, an alternative explanation for the excess of fast and ultra-fast rotators  in $\alpha$ Per, M~34, and M~11.

Two rotational tracks, with initial conditions representative of slow ($\Omega_e(t_0)=\Omega_c(t_0)= 3~\Omega_\odot$) and fast rotators ($\Omega_e(t_0)=\Omega_c(t_0)= 25~\Omega_\odot$), as calculated with the DZM and prescriptions \eqref{eq:2val} or \eqref{eq:powl} for $\tau_c$, are compared with the observed distributions and the solar-age angular velocity in Figures \ref{fig:ev_2val} and \ref{fig:ev_plaw}, respectively.
Assigning a constant value to $\tau_c$ for the whole rotational evolution results in slow and fast tracks that intersect with each other at ages later than about $1$ Gyr (see Figure \ref{fig:ev_2val}).
This is another consequence of the high percentage of fast rotators in $\alpha$ Per, which imposes a fast angular momentum transfer from the core to the envelope just after the ZAMS.
This diminishes dramatically the core angular momentum reservoir available for later stages, producing a faster spin-down for fast rotators.
On the other hand, the power-law prescription for  $\tau_c$ leads to a remarkably greater internal differential rotation for the fast rotators and a somewhat lower internal differential rotation for the slow rotators than the two-value prescription.

The solar constraint (equation \ref{eq:sun_con}) is  reasonably matched by the power-law prescription.
The large differential rotation developed by fast rotators is due to the relatively large value of $\tau_c$, as previously noted.
In contrast, the poor matching found for the slow rotators in the two-value case may seem to be in contradiction with previous results (e.g. \citealt{Bouvier:2008b}).  
Note, however, that in the present work an attempt is made to fit the rotation period distributions, not just their percentiles. 
A value of $\tau_c \sim 130$ Myr, which is the result of our best-fitting procedure (and as such is the outcome of taking into account all the period distributions plus the solar constraint), turns out to be too large to recover the almost rigid rotation at the age of the Sun (cf. also Figure \ref{fig:tauc_plaw} for the power-law case). 
The slightly lower value of $K_w$ with respect to the power-law case (see Table \ref{tab:results}), moreover, explains the excess of rotation at $t_\odot$.
Indeed, this disagreement is partly relieved if the $\alpha$ Per distribution, which is likely prone to observational biases (see Sect. \ref{sec:disc}), is excluded from the fitting procedure.  
A run performed for comparison purposes without the $\alpha$ Per constraint in the two-value case gives KS probabilities above $20 \, \%$ for the intermediate age clusters (i.e. M~50 to M~11) and different values of the best fitting parameters, namely $K_w \sim 6.3 \cdot 10^{47}$ g cm$^2$ s and $\tau_c \sim 80$ Myr.
A similar test for the power-law case results in negligible differences both in KS probabilities and in best-fit parameters, consistently with a moderate sensitivity to the $\alpha$ Per constraint.

In conclusion, the power-law prescription is more suited to fit the shape of the rotational distributions, while the two-value models better reproduces the basic features of the two extreme populations (i.e. the slow and fast rotators). 
The two prescriptions also show different sensitivity to peculiar characteristics (either real or due to biases) of the distributions.
Clearly, none of the two models is able to completely describe rotational evolution.

\section{Discussion}
\label{sec:disc}

We modelled the evolution of rotation period distributions from the PMS to the solar age, using the available observations to determine the best-fitting parameters of a modified version of the DZM.
We aimed at including in the analysis the whole statistical information contained in the  observed distributions as available so far; at the same time, we tested the possibility of including in the DZM some kind of dependence of the coupling timescale on the rotational state of the star. 
A numerical procedure was implemented to evolve forwards in time the solar analogues' subsample of the ONC period distribution, used as our initial condition.
The agreement of the synthetic distributions with the observed ones was evaluated by means of the KS test.
This procedure was used to select the values of model parameters ensuring the best fitting of the observational constraints.

A prescription for the dependence of $\tau_c$ on the rotation of the star is required to fit the evolution of the whole period distributions.
It is already known, in fact, that the slow and fast percentiles of the period distributions require significantly different values of $\tau_c$ \citep[e.g.][]{Bouvier:2008b}. 
In formulating these prescriptions (see equations \ref{eq:2val} and \ref{eq:powl}) we relied on basic physical intuition, keeping the number of free parameters to a minimum.
Equation \eqref{eq:2val} is the simplest way to mimic the dichotomy in $\tau_c$ between slow and fast rotators  as suggested by earlier works.
It implies that different $\tau_c$ values are set at early stages by some mechanism that quickly distinguishes between two separate rotation regimes, but is poorly sensitive to internal differential rotation. 
A different magnetic field configuration within the stellar interior might account for such a behaviour (see \citealt{Spruit:1999}; \citealt{Spada_ea:2010}).
Equation \eqref{eq:powl}, on the other hand, assumes that the available content of rotational kinetic energy, as measured in the DZM by the amount of differential rotation, can trigger angular momentum transport processes with an efficiency parameterised by the exponent $\alpha$. 
Other formulations, with a power law dependence of $\tau_c$ on the surface angular velocity $\Omega_e(t)$, with or without saturation mechanisms like that in equation \eqref{eq:wind}, proved to be completely unsuccessful.

A major limitation of this work comes from the uncertainty in the ages of the clusters \citep[e.g.][and references therein]{Mayne_Naylor:2008}.
This is particularly relevant for the very young clusters and for the stellar associations, which are also more likely to be prone to remarkable internal age scattering.

 During the PMS, the fitting is dependent essentially on $\tau_\mathrm{disc}$ and on the radius contraction as predicted by the evolutionary model. 
Considering the uncertainties in age and the crude approximation on the role of the circumstellar disc, the fitting can be considered quite satisfactory in this part of the evolution.
After the ZAMS, when the rotational evolution becomes very sensitive to angular momentum transfer from the core to the envelope and to wind-braking, the observed CDFs do not show a regular evolution as expected from the DZM assumptions.
The $\alpha$ Per CDF, in particular, shows a marked (ultra) fast-rotators excess, which is difficult to reconcile with the CDF of younger systems, either assuming a two-value or a power-law prescription for $\tau_c$ in the DZM.
This distribution feature is not seen in any other cluster or association, except perhaps M~34 and M~11.
The consequences on the model fitting have been discussed in Sect.\,\ref{sec:results}.

Obviously, the reasons for such a discrepancy could be: 
a) the DZM is inadequate or contains oversimplified assumptions; 
b) the stellar rotation period distributions in open clusters and young associations may be affected by characteristics of the (parent) cluster as a whole, particularly those leading to the evaporation of circumstellar discs by UV radiation and wind from neighbour high-mass stars \citep[e.g.][]{Hernandez_ea:2008, Guarcello_ea:2010}; 
c) the observed period distributions are incomplete and/or biased. 

Indeed, our model does not include differences in the rotational history of clusters' stars  that may arise from the presence of high mass stars, multiple or prolonged star formation events, the compactness and richness of the clusters, and the stars' dispersal into the field.
Therefore, despite the DZM may contain the essential physics to describe the stellar angular momentum evolution, a satisfactory modelling may require taking into account the peculiarities of the cluster to which the stars belong. 
In the case of $\alpha$ Per, for instance, high mass stars might have influenced the rotational history of a fraction of low mass stars by evaporating their circumstellar discs at an early stage.
Without an efficient disc-locking, these stars may have reached the ZAMS  (first 30~Myr) with a very high rotation and have had no time to spin-down significantly in the subsequent $\sim 40$~ Myr.

Other possible sources of uncertainties in the period distribution include clusters contamination by field stars, the presence of close binaries whose high rotation rate is maintained by the tidal synchronisation between the components, and observational biases. 
Concerning these latter, photometric monitoring with a limited temporal extension (generally of $10$--$15$ d as in the cases of M~11, M~34, M~35, M~50, NGC~2362 and of numerous members of $\alpha$ Per) prevents us from detecting the slow rotation periods, making the derived distributions incomplete.
Moreover, for MS stars with $P_\mathrm{rot} \gtrsim 10$ d the amplitude of the optical light curve is very small ($< 0.001 - 0.01$ mag), so their rotational modulation cannot be measured by ground-based photometry \citep[see, e.g., ][]{Radick_ea:1998, Messina_ea:2003}.
The absence of rotators with periods longer than $\sim 10$ d in the observed distributions shown in Figures \ref{fig:pks_2val} and \ref{fig:pks_plaw} can therefore be due to  such limitations, although this is unlikely to hamper the analysis to a significant extent as the percentage of rotators with $P_\mathrm{rot} \gtrsim 10$ is predicted to be rather low at young ages. 
On the other hand, a poor sampling from ground-based observations (for example, $1$--$2$ d in the case of young associations considered in our analysis) makes it hard to detect the ultra-fast rotators, making the derived period distributions incomplete also on the fast side.
Despite high-precision CCD photometry and multi-site monitoring campaigns may overcome most of these difficulties, the  data available to date are still  rather inhomogeneous and far from being complete.
Most stars, especially in MS, do not show a permanent periodic modulation, and therefore deriving a rotation period is sometimes a matter of chance, unless stars are observed for several seasons \citep[see, e.g.,][]{Parihar_ea:2009}.
Often a beat period is detected and prolonged monitoring is required to remove aliasing effects.  
The effect of large spotted areas localised on opposite stellar hemispheres, producing a rotational modulation with half the true period, should also be taken into account \citep[see, e.g.,][]{Cameron_ea:2009}.

The dispersal of clusters' stars into the field may add other selection effects. In order to assemble a complete sample one should, in principle, collect data for all stars in each cluster as well as data for all cluster's stars that have been dispersed into the field, a task which is far from being accomplished.
Regarding $\alpha$ Per, which is particularly important to constrain the rotational evolution just after the ZAMS, an alternative explanation for the presence of a high percentage of fast and ultra-fast rotators is therefore that the dataset available may miss information about a significant fraction of slower rotators, either because of observational biases or because of an intrinsic incompleteness of the sample due to stars' dispersal into the field.
We note in passing that the fraction of ultra-fast rotators (i.e., $P_\mathrm{rot} \lesssim 2$ d) in the CDF of AB~Dor / Pleiades was reduced from $\sim0.2$ to the current $\sim0.12$ by the data recently acquired by \cite{Hartman_ea:2010}.  
The same reasoning may also apply to $\alpha$ Per, M\,50, M\,35,  M~34 and M~11.

\begin{figure}
\begin{center}
\includegraphics[width=0.45\textwidth]{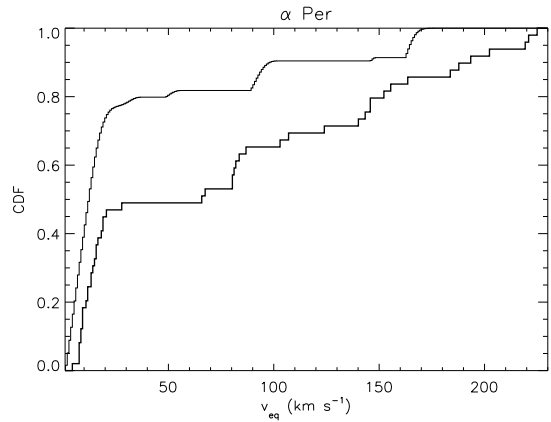}
\caption{Cumulative distributions for the equatorial rotational velocities derived from the $v \sin i$ data (thin line) and from the rotation period data (thick line) in $\alpha$ Per.}
\label{fig:aperbias}
\end{center}
\end{figure}

Some indication of possible observational biases in the alpha Per periods can be derived from a comparison with $v \sin i$ measurements. 
Using the \citet{Gaige:1993} inversion procedures for a random orientation of rotational axes and $v \sin i$ data from \cite{Stauffer_ea:1985, Stauffer_ea:1989}, we have derived the cumulative distribution of equatorial velocities for alpha Per stars in our mass range. Such procedures have been applied, for instance, by \citet{Queloz_ea:1998} in an analysis of rotational velocities in the Pleiades. 
We have then compared such a distribution with the cumulative distribution of equatorial velocities derived from the period measurements assuming that the radius of all stars in our mass range is approximately $1\, R_\odot$ (Fig. \ref{fig:aperbias}).

For the slowest rotators, \cite{Stauffer_ea:1985, Stauffer_ea:1989} are able to derive only an upper limit of $v \sin i$, which for the stars in our mass range is $\sim 10$ km/s. 
This implies that the cumulative distribution of recovered equatorial velocities is rather uncertain at the lower end of the velocity range. 
Numerical experiments with several assumptions on the $v \sin i$ distribution below $10$ km/s have shown, however, that above $v_{\rm eq} \sim 15$ km/s the derived cumulative distribution function is unaffected by any assumption made on the $v \sin i$ distribution below $10$ km/s. 
On the other hand, the expected differences in stellar radii in our mass range have negligible effects on the cumulative distribution of equatorial velocities derived from the period measurements. 
Therefore, a meaningful comparison can be done between the fractions of stars with rotational velocities lower than, say, $30$ km/s in both distributions, which are approximately $80\, \%$ according to the $v \sin i$ measurements and $50\, \%$ according to the rotational period measurements. 
Furthermore, from the \cite{Chandra:1950} relationships (the overline denotes average)
$$\overline{v \sin i} = \frac{\pi}{4} \bar{ v}$$ 
and 
$$\sigma^2= \overline{(v-\bar v)^2} = \frac{3}{2} \left[ \overline{(v \sin i)^2} - \frac{16}{\pi^2} \left(\overline{v \sin i} \right)^2 \right]$$
we derive $\bar v = 35.5$ km/s, and $\sigma^2 =2800$ (km/s)$^2$, while the rotational velocities inferred from the period measurements have $\bar v = 63.9$ km/s and $\sigma^2=3700$ (km/s)$^2$. 
Finally, the KS test applied to these cumulative distribution functions gives a probability of only $0.008$ that they are drawn from the same distribution. 
We conclude that there is a convincing evidence of a bias in the measured period distribution in $\alpha$ Per towards an excess of fast rotators. 
Having carried out such a test on the problematic period distribution of $\alpha$ Per, a similar comparison for other clusters in our sample would be desirable but it is outside the scope of this paper and is deferred to future work.

The DZM with a power-law prescription for $\tau_c$ is found quite effective in describing the rotational evolution from AB~Dor / Pleiades to M~37. 
Between these two, however, it predicts more rotators with  $P_\mathrm{rot} \gtrsim 5$ d than observed in M~50 and M~35.  
For M~34 (which seems to have also a higher fraction of fast rotators than predicted) and M~11 the number of available periods seems too low to draw any definitive conclusion.
Assuming that the DZM contains the essential physics to describe the evolution from AB~Dor / Pleiades to M~37 and that the age estimate is not too poor, one may reach the conclusion that a significant fraction of rotators with $P_\mathrm{rot} \gtrsim 5$~d are missing in M~50 and M~35.
This conclusion is also supported by a crude comparison between the slower rotator tail in the observed CDFs, from which M~50 and M~35 would appear rotationally older than M~37.
At ages between the ZAMS and AB~Dor /Pleiades, the DZM with a power-law prescription for $\tau_c$ is unable to describe the high percentage of fast and ultra-fast rotators in $\alpha$ Per, as discussed above, and is not able to fit the Tuc-Hor/Col/Car, despite the shape of the observed and fitted CDFs' are not too different.
For these latter associations, however, it should be noted that, besides the issue of completeness discussed above, the modelling is particular sensitive to age uncertainties in this evolution phase and to the $\alpha$ Per constraint at 70~Myr.

The fit using the two-value prescription for $\tau_c$ is more affected by the high percentage of (ultra) fast rotators in $\alpha$ Per than the power-law prescription. 
This prevents from obtaining a satisfactory fit for older systems, where such a high percentage of fast and ultra-fast rotators is not present. 

Only for M~34, which also seems to show a high percentage of fast rotators, $P_{\rm KS,i}$ is close to unity. 
Also for Tuc-Hor/Col/Car the two-value prescription for $\tau_c$ produces a synthetic CDF which is slightly higher than the power-law prescription for the faster rotators partly because of the constraint imposed by $\alpha$ Per.

\section{Conclusions}

We studied the rotational evolution of solar-mass stars from PMS to the solar age, using the most complete sample of observations on rotation period distributions available to date, filling the gap between $\sim 6$ and $70$ Myr with recently acquired data and implementing a novel prescription on the internal rotational coupling in the framework of the DZM.

Although the completeness and accuracy of observational data have considerably improved in recent years, possible selection effects (e.g. scarce sensitivity to rotational modulation with $P_\mathrm{rot} \gtrsim 10$ d, limits for detection of fast or slow rotation periods due to insufficient time sampling or extension of the observations), as well as the uncertainties in clusters' ages, are still severe limitations for the use of the period distributions to study stellar rotational evolution.   
Nevertheless, we investigated the dependence of the core-envelope coupling timescale $\tau_c$ on stellar rotation. 

After the ZAMS, a power-law dependence of $\tau_c$ on the stellar rotational velocity produces CDFs whose shapes are more similar to observations than the two-value prescription and gives an excellent fit for AB Dor / Pleiades and M\,37. 
While M\,34 and M\,11 have too few period measurements to draw firm conclusions, the discrepancies in $\alpha$ Per, M\,50, and M\,35 could be due either to observational biases or to environmental effects, such as those causing early destruction of circumstellar discs.

Comparing the results obtained with the two-value prescription for $\tau_c$ with those obtained with the power-law, it appears that the former better reproduce the extrema while the latter the shape of the observed period distributions. 
This conclusion may be a useful guide to formulate more sophisticated models of rotational evolution in the future. 

Part of the discrepancy between the models and the observations here may also arise from the assumption of a single disk lifetime ($\tau_{\rm disc}=5.8$ Myr). 
Assuming instead a distribution of disk lifetimes would yield more fast rotators at a given age and a broader rotation period distribution at the ZAMS, but including this feature in the model is beyond the scope of this paper.

To put our conclusions on a firm ground, however, more observations of clusters and associations in the age range $30 - 100$~Myr are required, together with an improvement of the model, e.g., by including correlations with the global properties of each cluster.

\section*{Acknowledgements}
The authors thank an anonymous referee for her/his valuable comments.
FS thanks Professor J. Bouvier for interesting discussion.
This research has made use of the ADS-CDS databases, operated at the CDS, Strasbourg, France.

\bibliographystyle{mn2e}


\label{lastpage}

\end{document}